# Investigating the Impact of SOLID Design Principles on Machine Learning Code Understanding

Raphael Cabral
rcabral@inf.puc-rio.br
PUC-Rio
Rio de Janeiro, Brazil

Marcos Kalinowski
PUC-Rio
Rio de Janeiro, Brazil
kalinowski@inf.puc-rio.br

Maria Teresa Baldassarre
University of Bari
Bari, Italy
mariateresa.baldassarre@uniba.it

Hugo Villamizar
PUC-Rio
Rio de Janeiro, Brazil
hvillamizar@inf.puc-rio.br

Tatiana Escovedo
PUC-Rio
Rio de Janeiro, Brazil
tatiana@inf.puc-rio.br

Hélio Lopes
PUC-Rio
Rio de Janeiro, Brazil
lopes@inf.puc-rio.br

## ABSTRACT

[Context] Applying design principles has long been acknowledged as beneficial for understanding and maintainability in traditional software projects. These benefits may similarly hold for Machine Learning (ML) projects, which involve iterative experimentation with data, models, and algorithms. However, ML components are often developed by data scientists with diverse educational backgrounds, potentially resulting in code that doesn't adhere to software design best practices. [Goal] In order to better understand this phenomenon, we investigated the impact of the SOLID design principles on ML code understanding. [Method] We conducted a controlled experiment with three independent trials involving 100 data scientists. We restructured real industrial ML code that did not use SOLID principles. Within each trial, one group was presented with the original ML code, while the other was presented with ML code incorporating SOLID principles. Participants of both groups were asked to analyze the code and fill out a questionnaire that included both open-ended and closed-ended questions on their understanding. [Results] The study results provide statistically significant evidence that the adoption of the SOLID design principles can improve code understanding within the realm of ML projects. [Conclusion] We put forward that software engineering design principles should be spread within the data science community and considered for enhancing the maintainability of ML code.

## 1 INTRODUCTION

Contemporary advances in Machine Learning (ML) and the availability of vast amounts of data have both given rise to the feasibility and practical relevance of incorporating ML components into software-intensive systems. ML is inherently driven by experimentation, requiring data scientists to explore data, algorithms, and models to find the most satisfying way of achieving their objectives [1]. Moreover, ML is also often used in the context of proofs of concept, which allow for iterative refinement and validation before implementation into a production environment. This may encourage quick deliveries over clean code.

Furthermore, data scientists in charge of developing these ML components may have a variety of educational backgrounds, such as economics, mathematics, and physics, and typically lack Software Engineering (SE) foundations [9]. Therefore, ML code often falls short of adhering to software development best practices, resulting in low-quality code that may pose challenges in terms of maintenance and long-term sustainability [22].

Despite code comprehension being studied for over 40 years [25], and the existence of evidence indicating that maintenance requires a considerable amount of work related to understanding code [5, 6, 10, 14, 26], we are not aware of studies extending code understanding investigations to the domain of ML code. This gap is particularly noteworthy because there seem to be several ways in which ML code could benefit from well-established SE practices. For instance, as data scientists often work on common ML components such as data pre-processing, model training, evaluation, and deployment, they could break down that code into reusable modules or functions, enhancing code reuse and saving time and effort.

In response to the dynamic environment of ML development and recognizing this identified research gap, in this paper, we investigate the impact of using SOLID design principles - which are well-known object-oriented design principles for writing clean code [13] - on ML code understanding capabilities of data scientists. More specifically, we conducted a controlled experiment with 100 data scientists from three different organizations that were divided into two groups. The control group was presented with ML code from a real industrial setting that did not incorporate SOLID principles, while the experimental group was presented with that same ML code restructured by applying the SOLID principles. Subsequently, the data scientists were tasked with analyzing the code and filling out a questionnaire that included closed-ended and open-ended questions related to their understanding of the code and their agreement with statements related to typical implications of applying the SOLID principles.

The results indicate that the adoption of each of the five SOLID design principles significantly improves ML code understanding. More specifically, the application of the principles was also perceived to lead to expected benefits related to having clearly distributed and defined ML code responsibilities, facilitating ML code extensions without substantially changing existing code, favoring low coupling and enabling substituting ML code elements, and resulting in proper segregation of interface operations, not forcing ML code to depend upon methods that it does not use.

## 2 BACKGROUND AND RELATED WORK

This section provides the background on SOLID principles and on challenges related to ML code comprehension that further motivate our investigation.

### 2.1 SOLID Principles

Several object-oriented design principles have emerged, such as SOLID [12] and GRASP [11], aiming at making code more understandable, flexible, and maintainable. In this paper, we focus on the five SOLID principles, which encapsulate fundamental design concepts. This choice was also motivated by the inclusion of these principles in best practices books for software developers and popular textbooks on software engineering for data science and machine learning (*e.g.*, [7, 13]), their consideration in academic investigations regarding design principles in general (*e.g.*, [3]), grey literature industrial sources (*e.g.*, [21]) and data science blog posts (*e.g.*, [15, 19]) anecdotally advocating for benefits of using SOLID within the ML context, and even online courses teaching "SOLID Principles for Machine Learning Engineers" (*e.g.*, [23]). Hence, while other design principles might also be of interest, the need to dig deeper into the use of the SOLID principles within ML is supported by multifaceted scientific and practical motivations. SOLID is an acronym that refers to five object-oriented principles that seek to make software products understandable, maintainable, and reusable. Hereafter, we briefly describe each of these principles based on [13].

The *Single Responsibility Principle* (SRP) advocates that each class only has responsibility for one part of the software's function; *i.e.* each class does just one thing, keeping the design as simple as possible, and future changes and updates easier and less disruptive. The *Open-Closed Principle* (OCP) emphasizes that classes should be open for extension but closed for modification. This means it is possible to extend a class's behavior without altering its source code. In that direction, the *Liskov Substitution Principle* (LSP) asserts that derived classes should be replaceable for their parent classes without changing the function's correctness. Like OCP, this aims to contribute to long-term maintainability by ensuring that extensions do not fundamentally alter existing functionality.

The *Interface Segregation Principle* (ISP) encourages the creation of multiple, client-specific interfaces rather than attempting to extend the use of an existing interface. *I.e.*, a client should never be forced to implement an interface that it doesn't use, or clients shouldn't be forced to depend on methods they do not use. The last one, called *Dependency Inversion Principle* (DIP), has the following premise: "Depend on abstractions, not implementations". The idea is that whenever it is necessary to couple to another class or module, the less stable one should depend on the more stable one. A stable class is one that tends to change rarely. The advantage is that if it doesn't change, it also won't propagate change to the implementation.

### 2.2 Challenges in ML Code Comprehension

According to a systematic mapping conducted by Wyrich *et al.*[25], the importance of code comprehension has been recognized for over 40 years by the scientific community. Several studies show that developers invest a considerable amount of their daily work in understanding code [5, 6, 10, 14, 26].

ML projects present unique challenges in terms of code comprehension. One of the primary factors contributing to these challenges is the inherently experimental nature of ML. Data scientists often work on interactive computational notebooks [18] that combine code, text, and execution results to handle data, train, and evaluate models iteratively. This emphasis on rapid prototyping and literate programming documents can encourage poor coding practices and results that are difficult to reproduce [16, 17].

In addition to the experimentation-driven challenges, data scientists have a variety of educational backgrounds, such as statistics, physics, mathematics, and economics [1]. While this diversity of expertise is valuable for tackling complex ML problems, it also poses challenges for code comprehension. Many of these team members may lack formal SE foundations, leading to code that may not adhere to established best practices. In this line, Kim *et al.* [8] state that it is difficult to find data scientists who combine analytical and SE skills. As a consequence, there have been studies reporting the prevalence of code smells in ML code [22] and indicating that ML code is commonly subject to technical debt and refactoring [20].

Despite these challenges and the importance of code understanding for effective maintenance, we found no studies related to ML code comprehension in the literature. In this paper, we provide an initial step to address this gap by investigating the impact of using the SOLID design principles on ML code understanding.

## 3 EXPERIMENTAL STUDY PLAN

We decided to conduct a controlled experiment to investigate the impact of specific factor levels (using or not using the SOLID design principles) on code understanding, in isolation from other confounding factors [24]. Hereafter, we detail the experimental study planning steps as suggested by Wohlin *et al.* [24] (Figure 1), as well as the study operation, data collection, and analysis procedures. While we assessed and mitigated threats to validity during the experimental study planning, we will discuss them in Section 6.

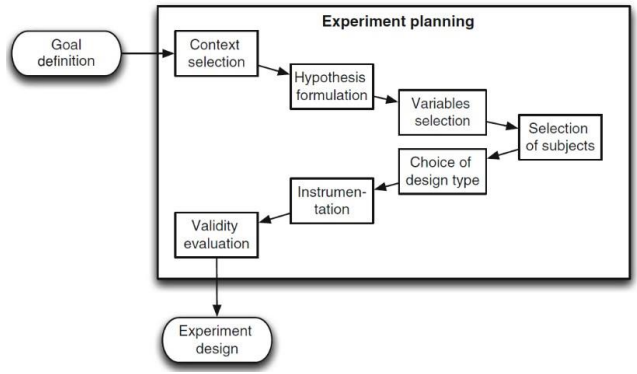

**Figure 1: Experiment planning steps [24]**

## 3.1 Goal Definition

The goal was defined using the Goal-Question-Metric (GQM) goal definition template proposed by Basili *et al.* [2] as follows: "Analyze the <application of SOLID design principles> for the purpose of <characterization> with respect to their <impact on ML code understanding> from the point of view of <data scientists> in the context of <industrial ML code>."

## 3.2 Context Selection

The selected context consists of an off-line experiment (*i.e.*, without direct participation of the researchers), where data scientists (students and professionals, as detailed and characterized later) were randomly assigned to two different treatments and asked questions related to their understanding of real industrial ML code.

To increase representativeness, we decided to use the ML code of a real ML-enabled system. The system was built to, based on an ML model prediction, emit alerts for oil refineries about the likelihood of emitting strong odors that could result in claims from the community. The ML code for this system was produced in Python using Jupyter Notebooks without employing SOLID design principles. It is noteworthy to mention that the system has been deployed and is currently in use in several oil refineries.

While part of a specific solution, the ML code sample is representative of code typically part of ML applications, including code for tasks such as data loading, preprocessing, ML model building, and ML model evaluation. The treatments involved receiving either the original ML code or the original ML with a restructuring using the SOLID design principles and answering the same questions.

## 3.3 Hypothesis Formulation

We formulated the following null hypothesis (H0) and alternative hypothesis (H1). We aim to reject the H0 by showing statistically significant differences in the levels of code understanding with an alpha value of 0.1.

- H0: There is no relationship between applying the SOLID Design Principles and ML code understanding.
- H1: Applying the SOLID Design Principles improves ML code understanding.

## 3.4 Variables Selection

The independent variable of interest (experimental factor) corresponds to the application of the SOLID design principles. It can receive as treatments the use or not use of SOLID Principles. Other independent variables captured during the subject characterization included experience with ML software development and SOLID.

The experimental study has two dependent variables. The first is the difficulty level of understanding of the source code perceived by the developer. Therefore, we prepared ordinal Likert scale questions on the difficulty in understanding the source code, which could assume the following values: 1 - Very Hard, 2 - Hard, 3 - Normal, 4 - Easy, 5 - Very Easy. The second is the level of agreement with statements related to the theoretical benefits of the SOLID principles. Again, we prepared ordinal Likert scale questions for these hypothesized benefits, which could assume the following values: 1 - Totally disagree, 2 - I disagree, 3 - I neither agree nor disagree, 4 - I agree, 5 - Totally agree.

## 3.5 Selection of Subjects

We used convenience sampling for our population of data scientists. The authors had access to data science graduate students from the University of Bari (Italy) and PUC-Rio (Brazil) and to data science professionals from SERPRO, a large-scale public IT company with more than 8,000 employees in Brazil.

We managed to have access to samples of 32 data science students from the University of Bari, 32 data science students from PUC-Rio, and 36 professional data scientists from the SERPRO, totaling 100 participants. Within these samples, we used probabilistic quotas to randomly assign the subjects to the experimental treatments in a balanced way. We characterized the subjects to allow us to apply experiment design principles, such as blocking, if needed.

## 3.6 Choice of Design Type

The goal was to investigate whether applying SOLID design principles produces ML code that is easier to understand. The dependent variables are the developer's level of understanding of the source code and the degree of agreement with statements related to the SOLID principles. Given this goal, we adopted a completely randomized one-factor with two treatments design [24]. The experiment setup uses the same object (the ML code of a real ML-enabled system) with two treatments (original and restructured, applying SOLID design principles) and assigns the subjects randomly to each treatment. The experimental tasks concerned participants who analyzed the source code and answered questions concerning difficulty in understanding and answering questions about their degree of agreement with statements related to the SOLID principles.

## 3.7 Instrumentation

We designed and independently peer-reviewed the instrumentation to ensure that it would provide the necessary means for appropriately collecting data for the experiment. It consisted of two versions of an online questionnaire (one for each treatment), implemented using Google Forms, divided into 4 basic components: a consent form, a participant characterization form, the substantive questions related to the difficulty in understanding code snippets and the levels of agreement with statements related to the SOLID principles, and a follow-up questionnaire. Both versions of the questionnaire are available in our online open science repository [4]. We detail each component of the instrumentation as follows.

*3.7.1 Consent form.* This form explains the research objectives, informs about their right to withdraw their participation at any time, and highlights the non-association of their name and e-mail address with the responses, ensuring that the research is conducted in accordance with ethical standards. It is noteworthy that all participants were volunteers.

*3.7.2 Participant Characterization Form.* This form aims to collect relevant demographic information that may influence the study's results. The form includes questions related to academic background, proficiency in the English language, experience with software development, specific experience with ML software development, and levels of expertise in specific software coding-related topics, such as object-oriented programming, SOLID principles, Design Patterns, and Python.

*3.7.3 Substantive questions related to the hypothesis.* These include a set of code snippets from the real ML code. Depending on the treatment, the original ones or the restructured ones (applying SOLID design principles). Both instrument versions can be found in the online repository [4]. Participants were required to analyze the code snippets and respond to closed-ended questions related to their perceived difficulty in understanding the code. Responses were registered using the five-point Likert scale: 1 - Very Hard, 2 - Hard, 3 - Normal, 4 - Easy, 5 - Very Easy. Additionally, participants are asked to provide feedback on their level of agreement with statements related to the SOLID principles, also on a five-point Likert scale: 1 - Totally disagree, 2 - I disagree, 3 - I neither agree nor disagree, 4 - I agree, 5 - Totally agree. Each closed-ended question had an optional open-ended question to provide justification for responses.

*3.7.4 Follow-up questionnaire.* Contains a closed-ended question regarding the participant's availability to answer further questions and two open-ended questions that provide an opportunity for feedback, allowing the participant to suggest improvements.

## 3.8 Operation and Data Collection

Volunteer subjects were selected as described in Section 3.5. Following the suggestion for controlled experiments, the choice of treatment applied to the participants was performed randomly [24]. The aim was to ensure that the two treatments were impartially assigned to the participants in order to avoid any bias. To achieve this randomization, participants were provided with a link to a website in which a JavaScript-implemented algorithm would randomly forward the participant to one of the two available questionnaires.

Data was collected from the participants' responses to the online questionnaires. The data collected during the experiment is also available in our online repository [4].

## 3.9 Analysis Procedures

To visually analyze the results, we used the frequency distributions of responses for the different treatments on the difficulties of source code understanding and levels of agreement with statements related to the SOLID principles. These frequencies were normalized and grouped by participant's origin and type of treatment applied.

To provide an overview of the distribution of responses provided by participants in the experiment, we calculated measures of central tendency, mean, median, and mode. Furthermore, to understand the distribution of the data, we calculated two measures of dispersion: the mean absolute deviation and the standard deviation.

For hypothesis testing, given the use of an ordinal Likert scale, it would not be possible to assume that the samples meet certain data distribution assumptions, such as a normal distribution, which led us to perform non-parametric hypothesis tests using Mann-Whitney. To measure the statistical effect size, we use Cohen's d, which is widely used for this purpose and easy to interpret.

## 4 EXPERIMENTAL STUDY RESULTS

In this section, we describe the participant characterization and present the experimental study results.

### 4.1 Participant Characterization

The participants were randomly assigned to the two treatments: SOLID (the SOLID restructured ML code) and unstructured (the original ML code). As described, participants were selected from three sources: University of BARI, PUC-Rio, and SERPRO. The distribution of participants for each treatment and source is presented in Table 1. It is possible to observe that the equal likelihood of being assigned to one treatment or the other led to almost balanced distributions of participants between treatments.

**Table 1: Participants by source and treatment**

| Treatment | SOLID | Unstructured |
|-----------|-------|--------------|
| BARI | 15 | 17 |
| PUC-Rio | 16 | 16 |
| SERPRO | 17 | 19 |
| Total | 48 | 52 |

The distribution of the level of education of participants per treatment and source is shown in Table 2, which also shows whether participants had academic degrees related to computer science or not. It is possible to observe that the random assignment led to some differences, which we do not believe to affect the results. In particular, slightly higher educational levels can be observed for the Unstructured treatment at the University of Bari and for the SOLID treatment at PUC-Rio, while the ones for the SERPRO company are comparable. It is also possible to observe a balance between participants with a computer science degree and from other areas. It is noteworthy, however, that the sources for subject selection may have led to a higher number of data scientists with a computer science background than in general. The data science graduate programs at both universities were part of the informatics departments, and the company was an IT company. We will further discuss this representativeness threat later.

**Table 2: Level of education and relation to computer science**

| Source | BARI | | PUC-Rio | | SERPRO | |
|---|---|---|---|---|---|---|
| Treatment | SOLID | Unstructured | SOLID | Unstructured | SOLID | Unstructured |
| Level of education | | | | | | |
| Bachelor's Degree. | 13 | 12 | 4 | 10 | 2 | 4 |
| Specialization. | 1 | 1 | 1 | 1 | 9 | 10 |
| Master's Degree. | 1 | 4 | 8 | 4 | 4 | 5 |
| Doctoral Degree. | 0 | 0 | 3 | 1 | 2 | 0 |
| Computer Science | 10 | 11 | 11 | 12 | 14 | 17 |
| Others | 5 | 6 | 5 | 4 | 3 | 2 |

The participants' experience with software development in general, as characterized across different treatments and sources, is presented in Table 3. Additionally, this table also shows the participants' years of experience in software development. Similarly, Table 4 shows the experience with specific ML-related software development. It is possible to observe that, in terms of experience, the control and experimental groups are comparable.

**Table 3: Experience with software development**

| Source | BARI | | PUC-Rio | | SERPRO | |
|---|---|---|---|---|---|---|
| Treatment | SOLID | Unstructured | SOLID | Unstructured | SOLID | Unstructured |
| Experience with software development | | | | | | |
| I have never developed software. | 0 | 0 | 1 | 0 | 0 | 0 |
| I have been developing for my own use. | 7 | 12 | 6 | 7 | 7 | 9 |
| I have been developing as team member, related to a course. | 14 | 16 | 8 | 7 | 10 | 12 |
| I have been developing as team member, in industry. | 1 | 5 | 12 | 16 | 17 | 17 |
| 0-2 years | 3 | 3 | 4 | 4 | 1 | 1 |
| 3-5 years | 9 | 10 | 4 | 6 | 1 | 4 |
| 6-10 years | 3 | 4 | 4 | 4 | 2 | 0 |
| More than 10 years | 0 | 0 | 4 | 2 | 13 | 14 |

**Table 4: Experience with ML Software Development**

| Source | BARI | | PUC-Rio | | SERPRO | |
|---|---|---|---|---|---|---|
| Treatment | SOLID | Unstructured | SOLID | Unstructured | SOLID | Unstructured |
| Experience with Machine Learning Software Development | | | | | | |
| I never developed ML software. | 3 | 4 | 1 | 0 | 0 | 1 |
| I developed ML software for my own use. | 4 | 6 | 6 | 8 | 10 | 7 |
| I developed ML software as a team member, related to a course. | 10 | 11 | 9 | 11 | 14 | 13 |
| I developed ML software as a team member in the industry. | 0 | 0 | 8 | 11 | 10 | 12 |
| 0-2 years | 12 | 15 | 10 | 11 | 6 | 8 |
| 3-5 years | 2 | 1 | 3 | 2 | 7 | 6 |
| 6-10 years | 1 | 1 | 1 | 3 | 2 | 5 |
| Greater than 10 years | 0 | 0 | 2 | 0 | 2 | 0 |

Finally, an overview of participants' background on software engineering and development topics related to this research, encompassing object-oriented programming, SOLID principles, and Python proficiency (among other collected data available in our online dataset), is presented in Table 5.

It is possible to observe that all of the participants had some experience with object-oriented programming and Python, with different experience levels of using SOLID principles. This data collectively provides valuable information about participants' backgrounds and competencies in software design aspects, offering context for interpreting their responses.

**Table 5: Experience with topics related to software engineering**

| Source | BARI | | PUC-Rio | | SERPRO | |
|---|---|---|---|---|---|---|
| Treatment | SOLID | Unstructured | SOLID | Unstructured | SOLID | Unstructured |
| O.O. Programming | | | | | | |
| I studied in a classroom or in a book. | 2 | | | | | 1 |
| I actively practiced in a classroom project. | 9 | 11 | 4 | 3 | 0 | 2 |
| I used it in one project in industry. | 2 | 2 | 1 | 7 | 2 | 4 |
| I used it in several projects in industry. | 2 | 3 | 10 | 5 | 15 | 12 |
| SOLID Principles | | | | | | |
| No experience. | 2 | 3 | 4 | 0 | 6 | 7 |
| I studied in a classroom or in a book. | 5 | 4 | 5 | 5 | 4 | 4 |
| I actively practiced in a classroom project. | 8 | 6 | 2 | 5 | 2 | 4 |
| I used it in one project in industry. | 0 | 2 | 3 | 5 | 0 | 1 |
| I used it in several projects in industry. | 0 | 2 | 2 | 1 | 5 | 3 |
| Python | | | | | | |
| I studied in a classroom or in a book. | 1 | 4 | 2 | 0 | 0 | 1 |
| I actively practiced in a classroom project. | 10 | 7 | 1 | 2 | 6 | 5 |
| I used it in one project in industry. | 2 | 3 | 7 | 2 | 2 | 2 |
| I used it in several projects in industry. | 2 | 3 | 6 | 12 | 9 | 11 |

## 4.2 Study Results

As described in our analysis procedures, we created graphs dedicated to analyzing the results of all closed-ended questions. These graphical representations provide a more visual perspective on the frequency of responses normalized by the type of treatment applied.

As an example, let's analyze the results obtained for the perception of the difficulty of understanding the source code for the SRP principle. The original piece of ML code had unstructured and sequential code for data loading, pre-processing, model building, model evaluation, and result plotting. The SRP restructured code had *DataLoader*, *PreProcessor*, *Model*, *Evaluator*, and *Plotter* abstract classes with concrete implementations and a *Controller* class receiving instances of those concrete implementations and orchestrating the execution through the defined abstract class interfaces. Having these classes with single responsibilities implemented, the final code looks like the snippet in Figure 2. It basically instantiates the concrete classes of interest, assigns them to their abstractions, and provides them to the controller.

The orchestration done by the controller, using the single responsibility classes is shown in Figure 3. It is possible to observe that it intuitively asks the *DataLoader* to load a dataset, the *PreProcessor* to preprocess the data (dividing it into preprocessed training and test input and output data), the *Model* to train itself (*fit* method) and to predict the output for the test input, the *Evaluator* to evaluate the predictions, and the *Plotter* to plot the results.

```
4   # Create DataLoader
5   dataLoader = DataLoaderFromWorkspace(ws,dataset_name)
6   # Create PreProcessor
7   preProcessor = EspecificPreprossessor()
8   #Create Model
9   model = LinearRegressionModel()
10  #Create Evaluator
11  evaluator = RegressionEvaluator()
12  #Create Plotter
13  plotter = MatPlotter()
14
15  controller = Controller(dataLoader,preProcessor,model,evaluator,plotter)
16  controller.run()
```

**Figure 2: Providing objects of single responsibility classes to a controller**

```
class Controller():
    """Class that orchestrates a model execution."""

    def __init__(self,dataLoader:DataLoader,preProcessor:PreProcessor,model:Model, evaluator:Evaluator, plotter:Plotter):
        self.__dataLoader = dataLoader
        self.__preProcessor = preProcessor
        self.__model = model
        self.__evaluator = evaluator
        self.__plotter = plotter

    def run(self):
        # Load Dataset
        dataset = self.__dataLoader.dataset()
        #Preprocessing
        x_train, x_test, y_train,y_test = self.__preProcessor.preprocess(dataset)
        #Model Training
        self.__model.fit(x_train,y_train)
        #Model Predict
        predictions = self.__model.predict(x_test)
        #Evaluation
        self.__evaluator.evaluate(y_test,predictions)
        self.__plotter.plot(y_test,predictions)
```

**Figure 3: Orchestration by the controller**

The close-ended questions (odd question numbers) asked to the participants of both treatments can be seen in Table 6. The open-ended justification questions (even question numbers) were intentionally omitted. The first question concerns the difficulty of understanding the source code, where the main difference between treatments was the application of the SRP principle as described. The results at the University of Bari, PUC-Rio, SERPRO, and the aggregated results are shown in Figures 4, 5, 6, and 7.

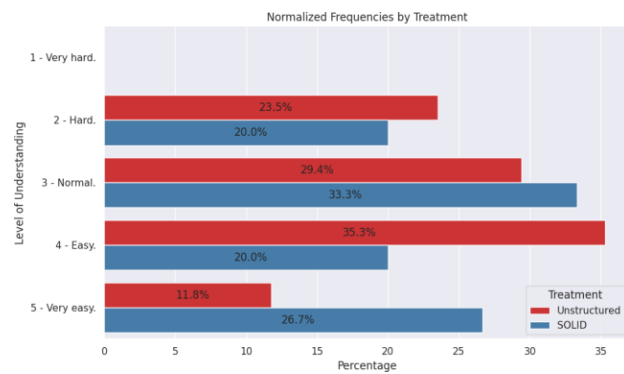

**Figure 4: Question 1 results at University of Bari: Perception about understanding the source code.**

While it is possible to observe higher perceived ease of understanding for the SOLID treatment in all cases (*e.g.*, higher prevalence of the "Very easy" and "easy" categories and a lower prevalence of the "Hard" category), the observed differences are statistically significant only for the SERPRO company and for the aggregated data (*cf.* Table 6). The subsequent close-ended questions assess the agreement with statements related to the definition and benefits of the SRP principle, which concern clearly distributed and defined responsibilities (Question 3), single responsibilities (Question 5), and single reasons for change (Question 7). For all of them, participants of the SOLID treatment agreed significantly more.

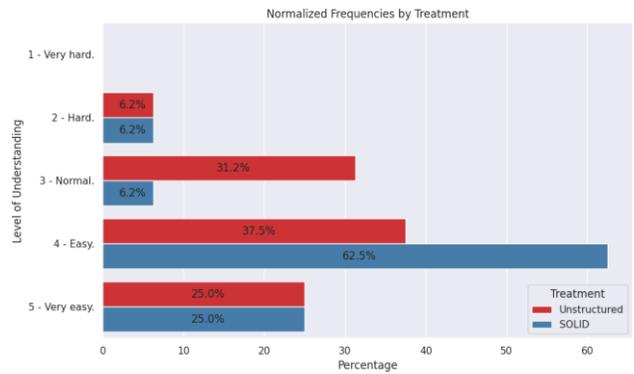

**Figure 5: Question 1 results at PUC-Rio: Perception about understanding the source code.**

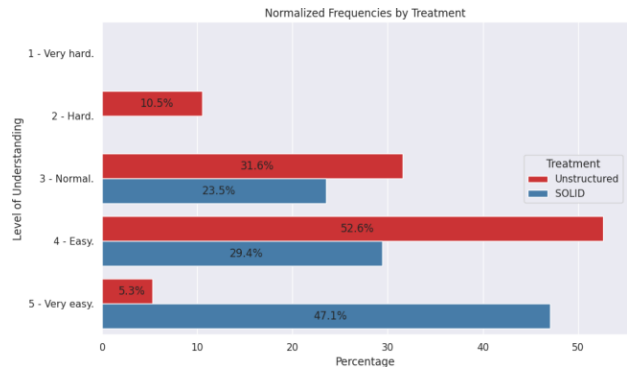

**Figure 6: Question 1 results at SERPRO: Perception about understanding the source code.**

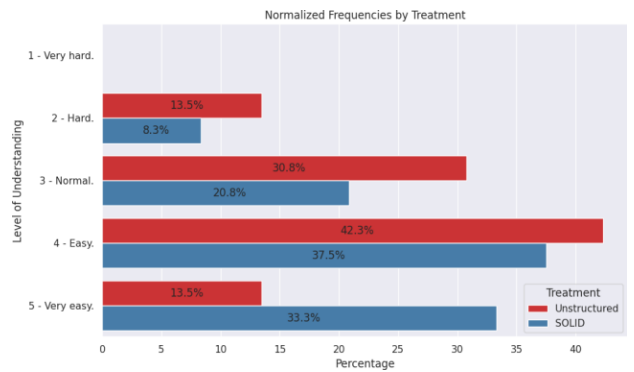

**Figure 7: The overall result for Question 1: Perception about understanding the source code.**

Due to space constraints, we will not go into the same depth for the other SOLID principles but briefly outline the code restructuring rationale. Details on the code can be obtained in the online instruments, and the 52 graphs for all 13 close-ended questions are also available [4]. For the OCP principle, the instrument introduces a scenario of a need to evolve the code to enable testing other regression algorithms. In the SOLID version, due to the OCP principle, the code is evolved by adding new ML model implementation classes extending the *Model* abstract class without having to change the behavior of existing code. Questions 9, 11, and 13 reflect the code understanding and the assessment of the agreement with the OCP expected benefits (ease to extend, not changing pre-existing code).

For the LSP and DIP principles, the instrument introduces an additional need. Given that, after the previous change, ML models can be built using several different regression algorithms, the plurality of regression models brings the need for the controller to be able to run a list of models. The LSP manifests in the code by having the abstract class *Model* defining the same interface as its child classes (strong behavioral subtyping). This allows models to be replaced within the list of models to be executed without changing the controller. The controller then uses the DIP principle to depend only on the more generic *Model* abstraction. Questions 15 to 21 capture these aspects related to the LSP and DIP principles.

Finally, for the ISP principle, the need to mix up regression and classification models for the same dataset is introduced. The SOLID ML code snippet introduces two segregated interfaces extending the *Evaluator* abstraction: *RegressionEvaluator* and *ClassificationEvaluator*. This is justified by regression and classification models typically being evaluated based on different metrics (*e.g.*, mean squared error and R2 score for regression, and accuracy, recall, precision, and f1-score for classification).

Table 6 contains the results for each experimental study trial and also the aggregated results combining data from all trials, allowing a comprehensive overview of the results on the effects of applying SOLID principles on the ease of understanding ML code and the expected SOLID principle benefits. It is noteworthy that all trials reflect exact replications of the experimental study, using the same instrument and random subject assignment, with balanced control and experimental groups. Statistically significant differences favoring the SOLID treatment are highlighted in green.

It is possible to observe that the experiment resulted in statistically significant differences for most questions in all trials. In particular, it is noteworthy that all questions had significant differences favoring the SOLID treatment when considering the results for the SERPRO's data science professionals (slightly larger sample with 36 participants) and for the aggregated results considering the 100 experimental study participants.

# 5 DISCUSSION

We discuss the obtained results for each SOLID principle based on the results shown in Table 6, focusing on the observed statistical significance and effect sizes.

## 5.1 SRP - Single Responsibility Principle

In the context of the SRP principle, the results show that the null hypothesis (H0) could not be rejected for the perception of the difficulty of understanding the source code (Question 1) for the trials at the University of Bari and PUC-Rio but could be rejected for the SERPRO trial and considering the aggregated results. Still, it is possible to observe a slightly higher perceived ease of understanding for the SOLID treatment in all trials (*cf.* Figures 4 and 5). Not achieving statistically significant differences between the University of Bari and PUC-Rio could be due to the slightly smaller sample size.

While a complete qualitative analysis of the open-ended questions is out of the scope of this paper, we used this data aiming at better understanding situations in which HO could not be rejected. At the University of Bari, three participants of the SOLID treatment considered the code hard to understand. Out of these three, two explained their answers. One mentioned "I'm quite new in the ML field. I've just studied some concepts for my personal projects" while the other one justified "I didn't understand the part of features and labels because I have no solid base of ML." Hence, their difficulties were apparently more closely related to specific ML-related implementations than to the code structure. At PUC-Rio, only one participant of the SOLID treatment considered the code hard to understand, justifying it with a "lack of practice with object-oriented programming". Another important aspect is that the original code was relatively small, potentially making the benefits of distributing responsibilities clearer for the subsequent maintenance scenarios. Indeed, many subjects of the Unstructured treatment also found the code easy to understand. The limited sample size within a single trial can have isolated cases of confounding factors leading to non-statistically significant results in some cases.

The SRP questions regarding clearly distributed and defined responsibilities (Question 3), single responsibilities (Question 5), and single reasons for change (Question 7) allowed rejecting H0, with significantly low p-values and high effect sizes. These results indicate that when it comes to specific aspects related to SRP, a significant and positive impact was perceived by the participants.

> **Finding 1:** Applying the SRP principle favors ML code understanding and leads to the perception of benefits related to having clearly distributed and defined ML code responsibilities, favoring future maintenance by having single responsibilities and single reasons for change.

## 5.2 OCP - Open-Closed Principle

For OCP, the null hypothesis (H0) could also not be rejected concerning the overall perception of the difficulty of understanding the related source code (Question 9) for the University of Bari and PUC-Rio. Nevertheless, a slightly higher perceived ease of understanding was observed for the SOLID treatment for all trials, even for the University of Bari and PUC-Rio (the difference can be observed in the graphs provided in the online material [4]). Not achieving statistically significant differences could be due to the smaller sample size. Indeed, it was possible to reject H0 for the SERPRO trial and for the aggregated results.

**Table 6: Consolidated results**

| | | SRP | | | | | | | | | | |
|---|---|---|---|---|---|---|---|---|---|---|---|---|
| Question | p-value | BARI Reject H0 | Effect Size | p-value | PUC-Rio Reject H0 | Effect Size | p-value | SERPRO Reject H0 | Effect Size | p-value | Overall Reject H0 | Effect Size |
| 1- What is your perception about understanding the source code? | 0.35 | No | | 0.2 | No | | 0.01 | Yes | 0.88 | 0.02 | Yes | 0.43 |
| 3- The responsibilities in the code above are clearly distributed and defined. What is your degree of agreement with the statement? | 0.01 | Yes | 1.15 | 0.01 | Yes | 2.25 | 0.01 | Yes | 1.46 | 0.01 | Yes | 1.56 |
| 5- The code block above has classes structured to have a single responsibility in the software, that is, they are specialized in a single subject. What is your degree of agreement with the statement? | 0.01 | Yes | 1.37 | 0.01 | Yes | 3.31 | 0.01 | Yes | 1.72 | 0.01 | Yes | 1.88 |
| 7- The code above is structured so that your classes have a single reason for change. What is your degree of agreement with the statement? | 0.01 | Yes | 0.98 | 0.01 | Yes | 1.86 | 0.01 | Yes | 1.46 | 0.01 | Yes | 1.42 |

| | | OCP | | | | | | | | | | |
|---|---|---|---|---|---|---|---|---|---|---|---|---|
| Question | p-value | BARI Reject H0 | Effect Size | p-value | PUC-Rio Reject H0 | Effect Size | p-value | SERPRO Reject H0 | Effect Size | p-value | Overall Reject H0 | Effect Size |
| 9- What is your perception about understanding the source code? | 0.19 | No | | 0.21 | No | | 0.02 | Yes | 0.72 | 0.01 | Yes | 0.48 |
| 11- It was easy to extend the behavior of the software with the addition of new models. What is your degree of agreement with the statement? | 0.04 | Yes | 0.50 | 0.01 | Yes | 1.1 | 0.01 | Yes | 0.65 | 0.01 | Yes | 0.75 |
| 13- Adding new models did not imply changing pre-existing code. What is your degree of agreement with the statement? | 0.07 | Yes | 0.52 | 0.01 | Yes | 1.44 | 0.01 | Yes | 1.29 | 0.01 | Yes | 0.82 |

| | | LSP and DIP | | | | | | | | | | |
|---|---|---|---|---|---|---|---|---|---|---|---|---|
| Question | p-value | BARI Reject H0 | Effect Size | p-value | PUC-Rio Reject H0 | Effect Size | p-value | SERPRO Reject H0 | Effect Size | p-value | Overall Reject H0 | Effect Size |
| 15- What is your perception about understanding the source code? | 0.02 | Yes | 0.78 | 0.01 | Yes | 0.73 | 0.02 | Yes | 0.71 | 0.01 | Yes | 0.74 |
| 17- The models can be replaced without changing the controller. What is your degree of agreement with the statement? | 0.09 | Yes | 0.43 | 0.27 | No | | 0.01 | Yes | 0.45 | 0.01 | Yes | 0.3 |
| 19- Changes in a model implementation do not imply changes in the controller. What is your degree of agreement with the statement? | 0.07 | Yes | 0.59 | 0.36 | No | | 0.01 | Yes | 1.07 | 0.01 | Yes | 0.54 |
| 21- The controller has loose coupling with model implementations. What is your degree of agreement with the statement? | 0.03 | Yes | 0.67 | 0.46 | No | | 0.01 | Yes | 1.32 | 0.01 | Yes | 0.65 |

| | | ISP | | | | | | | | | | |
|---|---|---|---|---|---|---|---|---|---|---|---|---|
| Question | p-value | BARI Reject H0 | Effect Size | p-value | PUC-Rio Reject H0 | Effect Size | p-value | SERPRO Reject H0 | Effect Size | p-value | Overall Reject H0 | Effect Size |
| 23- What is your perception about understanding the source code? | 0.06 | Yes | 0.55 | 0.04 | Yes | 0.72 | 0.04 | Yes | 0.48 | 0.01 | Yes | 0.55 |
| 25- The operations in the evaluator implementations are properly segregated for evaluation of classification and regression algorithms. What is your degree of agreement with the statement? | 0.01 | Yes | 0.92 | 0.01 | Yes | 2.14 | 0.01 | Yes | 3 | 0.01 | Yes | 1.82 |

Looking at the qualitative data at the University of Bari, two participants of the SOLID treatment considered the code difficult to understand, potentially leading to the lack of statistical significance in the differences. Out of these two, one provided an explanation: "I don't know these functions well," apparently referring to difficulties related to the new specific regression ML algorithms added as part of the OCP scenario and not to the code structure. At PUC-Rio, all participants perceived the code as easy or very easy to understand. Still, the difference favoring the SOLID treatment for PUC-Rio was not statistically significant.

The OCP questions regarding ease of extension (Question 11) and not implying changing existing code (Question 13) allowed rejecting H0, with significantly low p-values and medium to high effect sizes. This indicates that the application of OCP had a positive impact on the ease of extending the software without modifying existing code.

---

**Finding 2:** Applying the OCP principle favors ML code understanding and leads to the perception of benefits related to facilitating ML code extensions without substantially changing existing code.

---

## 5.3 LSP and DIP - Liskov Substitution Principle and Dependency Inversion Principle

For the LSP and DIP principles, the null hypothesis (H0) could be rejected with statistical significance and medium to high effect sizes concerning the overall perception of the difficulty of understanding the related source code (Question 9), indicating that the application of the principles made it easier to understand the ML code.

The specific questions related to model substitution (Questions 17 and 19) and low coupling (Question 21) also suggest that these principles had a significant and positive impact on the participants' perception regarding these aspects. An exception was observed for the trial at PUC-Rio, where these questions, while having answers slightly favoring the SOLID treatment [4], did not allow rejecting H0. Digging into the qualitative data allowed us to observe that for Question 17, out of the participants of the SOLID treatment at PUC-Rio, only one participant strongly disagreed, and one disagreed that models could be replaced without changing the controller, potentially leading to the lack of statistically significant differences within this trial. In their justifications, both argued that the controller had to be modified to support a list of models. Hence, they slightly misinterpreted the question, which referred to the ability to replace models within the list of models handled by the improved controller. This same misunderstanding led the one who strongly disagreed to also disagree with Questions 19 and 21, with which the participant who disagreed with Question 17 agreed, observing that changes in the implementation of a model would not affect the controller.

---

**Finding 3:** Applying the LSP and DIP principles favors ML code understanding and leads to the perception of benefits related to the flexibility of substituting ML code elements and favoring low coupling.

---

### 5.4 ISP - Interface Segregation Principle

For the ISP principle, the results show that the null hypothesis (H0) could be rejected with statistical significance and medium to high effect sizes for all questions related to this principle. This indicates that the application of ISP had a significant impact on the ease of understanding the ML source code (Question 23). Furthermore, it indicated that the ISP had a significantly positive impact on the participants' perception of proper segregation of operations, in this case, related to establishing separate evaluation interfaces for classification and regression algorithms.

---

**Finding 4:** Applying the ISP principle favors ML code understanding and leads to the perception of proper segregation of interface operations, not forcing ML code to depend upon methods that it does not use.

---

### 5.5 Implications

In summary, the consolidated analysis of the results with the combination of all trials indicates that the application of SOLID principles has a significant positive impact on ML code understanding and leads to the perception of several other benefits related to the ML system's maintainability. These results provide empirical evidence allowing to assert that the application of these principles is beneficial for improving ML code understanding and maintainability.

Implications for researchers include having evidence to ground themselves when referring to the need and benefits of applying software engineering design best practices within the ML context. Of course, this study can also foster additional experimental replications, potentially varying the experimental objects and involving larger and more diverse samples. Furthermore, additional studies are needed to investigate the effects of applying many other software engineering practices within the ML context.

Implications for data science practitioners include the need to familiarize themselves with software engineering design best practices in order to write more robust ML code. For educators, the results indicate that, as society becomes more and more dependent on ML-enabled systems, data science curricula should include software engineering skills to ensure that these systems and their ML components are built with quality and are easy to maintain.

## 6 THREATS TO VALIDITY

In this section, we describe the threats to validity faced by this research and the mitigation actions taken to control them within our possibilities. We organize the threats according to the categories described by Wohlin *et al.* [24].

### 6.1 Internal Validity

Threats to internal validity are trial-related influences that can affect the independent variable with respect to causality [24]. The instrument was applied without any assistance from the researchers. Hence, participants were not monitored during their activities, and some of them could have conducted the task with less attention or within environments facing interruptions, which may have affected their code understanding. Participants were volunteers, and we had no way to control this threat. However, all participants completed the tasks until the end and provided justification for at least some of the open-ended questions, which leads us to assume that they participated with attention.

Furthermore, before running the experimental trials, we conducted a pilot study with four participants, which allowed us to observe an average effort of 20 to 30 minutes. The pilot allowed us to improve the instrument with some simplifications (removing two questions that were considered redundant) and minor adjustments (*e.g.*, adjusting the size of the code, some snippets to improve readability, and adding definitions for some concepts). We considered the effort reasonable for retaining the attention of the participants.

### 6.2 External Validity

Threats to external validity are conditions that limit our ability to generalize the results of our experiment to industrial practice [24]. Two threats to external validity involve the representativeness of our subjects and the representativeness of our experimental object (i.e., the ML code used as a basis for the experiment) and tasks.

With respect to the subjects, we aimed to mitigate this threat by carefully characterizing them, allowing us to discuss their representativeness. Indeed, while we had data scientists with varying levels of experience in all three samples, 75% of our sample received education within the area of computer science. This does not properly reflect the observation of widely varying educational backgrounds reported in the literature [1, 8]. An explanation for this is that the data science graduate programs at both universities were part of the informatics departments and that the company was an IT company. Nevertheless, applying blocking allowed us to observe that an improvement in the ease of ML code understanding when SOLID principles are applied could also be observed for data scientists without a computer science degree. Due to the small sample size within this group not favoring significance testing or effect size calculations, we did not include this analysis in the paper. Indeed, we relied on convenience sampling to recruit the best sample we were able to in order to make this investigation happen and carefully describe its limitations. To further address this threat, we call for additional external replications with more diverse subjects.

Regarding the experimental object, to improve representativeness, we used the real ML code of a solution deployed at an industrial partner of PUC-Rio. This ML code was developed using Jupyter notebooks and without applying SOLID principles. For the SOLID treatment, to avoid any confounding factors, we strictly redesigned the code by applying the principles. It is noteworthy that the authors reviewed the code and that they included experts in software design. To assess the code in a maintenance context, we also elicited typical evolution scenarios that would make sense for the ML context based on our experience with delivering ML-enabled systems to industrial partners. These scenarios involved implementing several ML algorithms, assessing a list of ML models, and assessing regression and classification algorithms. All of these are commonly discussed tasks within the data science domain. Despite the care, we also call for additional replications involving other experimental objects and tasks.

### 6.3 Construct Validity

Construct validity concerns how well the treatments and outcome measurements of the experimental design reflect the causes and effects being assessed. We used real ML code as the basis for designing the experimental object. For one treatment, we used the ML code as it was, while for the other, we strictly applied the SOLID principles. Hence, this was the only difference in the treatments, and therefore, we believe that using the SOLID principles (or not) properly reflects the cause for observed differences (measured using the exact same questions and scales).

To avoid hypothesis guessing, we only informed participants that the study aimed to understand the impact of design principles on understanding ML code. They did not know which treatment they were randomly assigned to. Also, their participation was on a volunteer basis, and we informed them that the research would be conducted anonymously, avoiding evaluation apprehension.

To assess whether confounding factors would appear during the experimental tasks, we conducted a pilot study, which allowed us to improve the instrument further. We used the random assignment experimental design practice and characterized participants to control other confounding factors, such as background and experience. Nevertheless, we still observed some isolated confounding factors taking place, like a few participants not properly understanding some of the questions. Based on the qualitative analyses of the open-ended justifications, we observed that these were rare and isolated cases, not severely affecting the results. The instrument for each treatment and the collected data are available online [4].

### 6.4 Conclusion Validity

Conclusion validity refers to the degree to which conclusions about the relationship among variables based on the data are correct from a statistical point of view [24]. Conclusion validity can be affected by the sample size. We recruited 100 data scientists from 3 different origins: two universities and one company. For the inferential statistics, we conservatively employed the Mann-Whitney test with an alpha value of 0.1. Mann-Whitney is a non-parametric statistical test that doesn't pose assumptions on the data distribution and that can be safely applied to ordinal scales. Furthermore, we observed mainly comparable scenarios for the different experimental trials, improving our confidence in the results.

## 7 CONCLUDING REMARKS

ML projects present unique challenges in terms of code understanding. Among the primary factors contributing to these challenges are the inherently experimental nature of ML and the variety of educational backgrounds of data scientists. Despite these challenges and the importance of code understanding for effective maintenance, we found no studies related to ML code comprehension.

In this paper, we take a step towards addressing this gap, investigating the impact of SOLID design principles on ML code understanding by data scientists. We conducted a controlled experiment involving 100 data scientists from three different organizations. The control group was presented with ML code from a real industrial setting that did not incorporate SOLID principles. The experimental group was presented with that same ML code restructured by applying the SOLID principles. Subsequently, the data scientists were tasked with analyzing the code and filling out a questionnaire related to their understanding of the code and their agreement with statements related to typical implications of applying the SOLID design principles.

The results indicate that the adoption of each of the five SOLID design principles can significantly facilitate ML code understanding. Moreover, the application of the principles was also perceived to lead to expected benefits related to applying the SOLID principles. These benefits include having clearly defined ML code responsibilities, facilitating ML code extensions without substantially changing existing code, enabling substitution of ML code elements, favoring low coupling, and ensuring proper segregation of interfaces.

In light of these findings, we put forward the importance of disseminating design principles, such as SOLID, to familiarize the data science community with software engineering design best practices. Furthermore, we advocate for their consideration in industrial contexts, as a means to more robust and maintainable ML code.

## ACKNOWLEDGMENTS

We want to thank each of the 100 data scientists from the University of Bari, PUC-Rio, and SERPRO who voluntarily participated in our study; without them, this research would not have been possible. Thanks also to the Brazilian Research Council - CNPq (grant 312827/2020-2) and the Brazilian Coordination for the Improvement of Higher Education Personnel - CAPES (Finance Code 001).